\documentclass[paper]{JHEP3}
\usepackage{graphicx,amssymb}
\usepackage{citesort}

\title{Method 2 at NLO}

\author{
John Campbell
\\
Argonne National Laboratory 
\\
Argonne, IL 60439 USA
\\
{\it email: johnmc@hep.anl.gov}
}

\author{
Joey Huston
\\
Department of Physics and Astronomy
\\
Michigan State University
\\
East Lansing, MI 48824 USA
\\
{\it email: huston@pa.msu.edu}
}

\preprint{ANL-HEP-PR-04-58, 
          MSUHEP-040525,
          hep-ph/0405276}

\abstract{

This note describes a theoretical study of $Wb\overline{b}$ and $Wjj$
final states at the Tevatron using the NLO program MCFM. We 
extensively study the effect of NLO corrections with respect to variations
of input parameters such as the minimum jet $p_T$ and the  choice of renormalization
and factorization scales. In particular, we examine
possible implications for the Method 2 QCD background subtraction
technique for $t\overline{t}$ production. 
}

\begin{document}

\section{Introduction}

Final states involving vector bosons accompanied by a heavy quark pair
and/or light quark or gluon jets serve both as an arena for precision
tests of QCD as well as backgrounds to both Standard Model and
non-Standard Model physics~(see for example
\cite{Abachi:1995jf}--\cite{Acosta:2001ct}). Thus, it is important to
utilize the most important theoretical tools available for computing
their cross sections. In Run I at the Tevatron, predictions were
obtained using a leading order calculation (VECBOS,~\cite{vecbos})
supplemented by a parton shower Monte Carlo (HERWIG,~\cite{herwig}).
VECBOS calculates the production of $W+n$~jets and $Z+m$~jets for $n
\leq 4$ and $m \leq 3$, at leading order in the strong coupling
$\alpha_s$.

In Run II, there are many programs capable of calculating these processes to a
much higher jet multiplicity than was available in Run
I~\cite{alpgen,madgraph,madevent,comphep,grace}. Their use at the Tevatron has
been explored in a series of workshops at Fermilab~\cite{workshops}. In addition,
there is now a universal interface (the Les Houches accord~\cite{lhaccord})
between matrix element and parton shower Monte Carlo programs. However, the
above-mentioned matrix element programs are still leading-order calculations and
thus have a large theoretical uncertainty. In addition, there is still some
ambiguity in dealing with the soft and collinear cutoffs of the matrix element
programs when matching to a parton shower Monte Carlo, and thus a risk of under or
double-counting contributions. There are reasonable prescriptions for managing the
cutoffs currently in use in CDF~\cite{mlm}; in addition, a rigorous prescription
(CKKW~\cite{ckkw}) that  removes most  of the soft/collinear  cutoff dependence in
the matrix element to Monte Carlo merging has recently been adopted for use at the
Tevatron by Steve Mrenna and Peter Richardson~\cite{mrenna}. 

      It's only at next-to-leading order (NLO), though, that the normalization of
a calculation can be taken seriously.  Once more, the theoretical predictions have
evolved since Run I, where calculations were available for $W/Z+n$~jet production
only for $n \leq 1$ (DYRAD,~\cite{dyrad}). The current state of  the art for
calculations of this type involves $W/Z$ plus a $b\overline{b}$ pair, or plus two
jets~\cite{mcfm}. MCFM provides a calculation to NLO of $W/Z+b\overline{b}$ or
$W/Z+jj$ final states, but at the partonic level only. Soft and collinear
singularities are cancelled between the one-loop and tree level diagrams; $b$
quark mass effects have still not been included in the NLO matrix element
calculation, though, so a cut must be applied to ensure that the $b$ and
$\overline{b}$ are well-separated\footnote{MCFM imposes $m_{b\bar b}>4m_b^2$ to
regulate the divergence and also to ensure that the cross-section vanishes below
the physical threshold.}. 

      It is worth noting that there have been recent advances in the calculation
of one-loop multi-partonic final states (e.g. $Wb\overline{b}j$, $Wjjj$
~\cite{durham})  in a semi-automated fashion, but actual programs should not be
expected for at least a year, and so will not be available for the first
publications from Run II. It is also worth noting that the MC@NLO
program~\cite{mcatnlo} has successfully incorporated NLO matrix elements for $WW$,
$WZ$, $ZZ$, $t\overline{t}$ and $b\overline{b}$ and Higgs in a parton shower Monte
Carlo framework (HERWIG). Thus, the result is a fully exclusive final state at the
hadron level. It is very time-consuming, though, to add each new process; luckily
$Wb\overline{b}$ and $Wjj$ remain fairly high on the priority list. 

      In this note, we will examine some  of the characteristics of  the
$Wb\overline{b}$  final states using  MCFM at  both LO and NLO. In particular, we
will emphasize the impact of these calculations on our understanding of $\it
Method~ 2$ (which we outline in Section 5), used by CDF in both Run I and Run II
to determine backgrounds to $t\overline{t}$ production. We will directly calculate
the ratio $Wb\overline{b}/Wjj$ and discuss the general features of  this ratio for
higher jet multiplicities.


\section{Topologies}

\begin{figure}
\begin{center}
\includegraphics[scale=1.0]{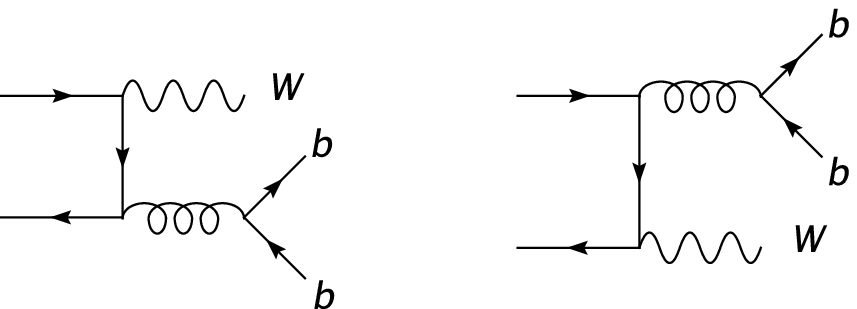}
\end{center}
\caption{ Leading order diagrams  for $Wb\overline{b}$ production.
\label{fig:wbb}}
\end{figure}

\begin{figure}
\begin{center}
\includegraphics[scale=1.0]{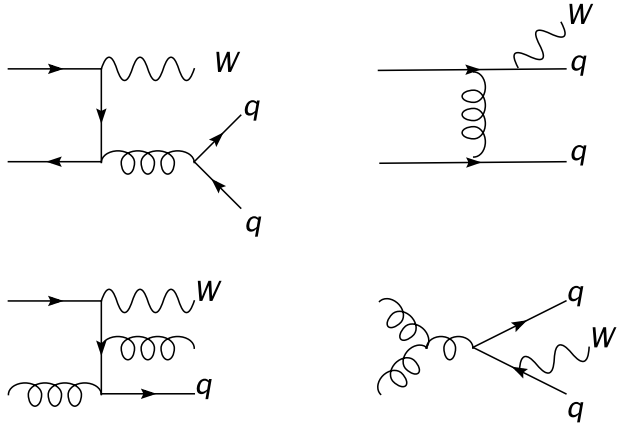}
\end{center}
\caption{A few of the  leading order diagrams  for $Wjj$ production.
\label{fig:w2}}
\end{figure}

      The topologies for $Wb\overline{b}$ production (see Figure~\ref{fig:wbb})
are much simpler than those for $W + 2$~jets (see Figure~\ref{fig:w2}). At LO,
there are basically two diagrams for $Wb\overline{b}$ production (involving only a
$q\overline{q}$ initial state) compared to literally hundreds of diagrams for the
production of $W + 2$~jets (involving both $q\overline{q}$, $gq$, and $gg$ initial
states).

\begin{figure}
\begin{center}
\includegraphics[scale=0.4]{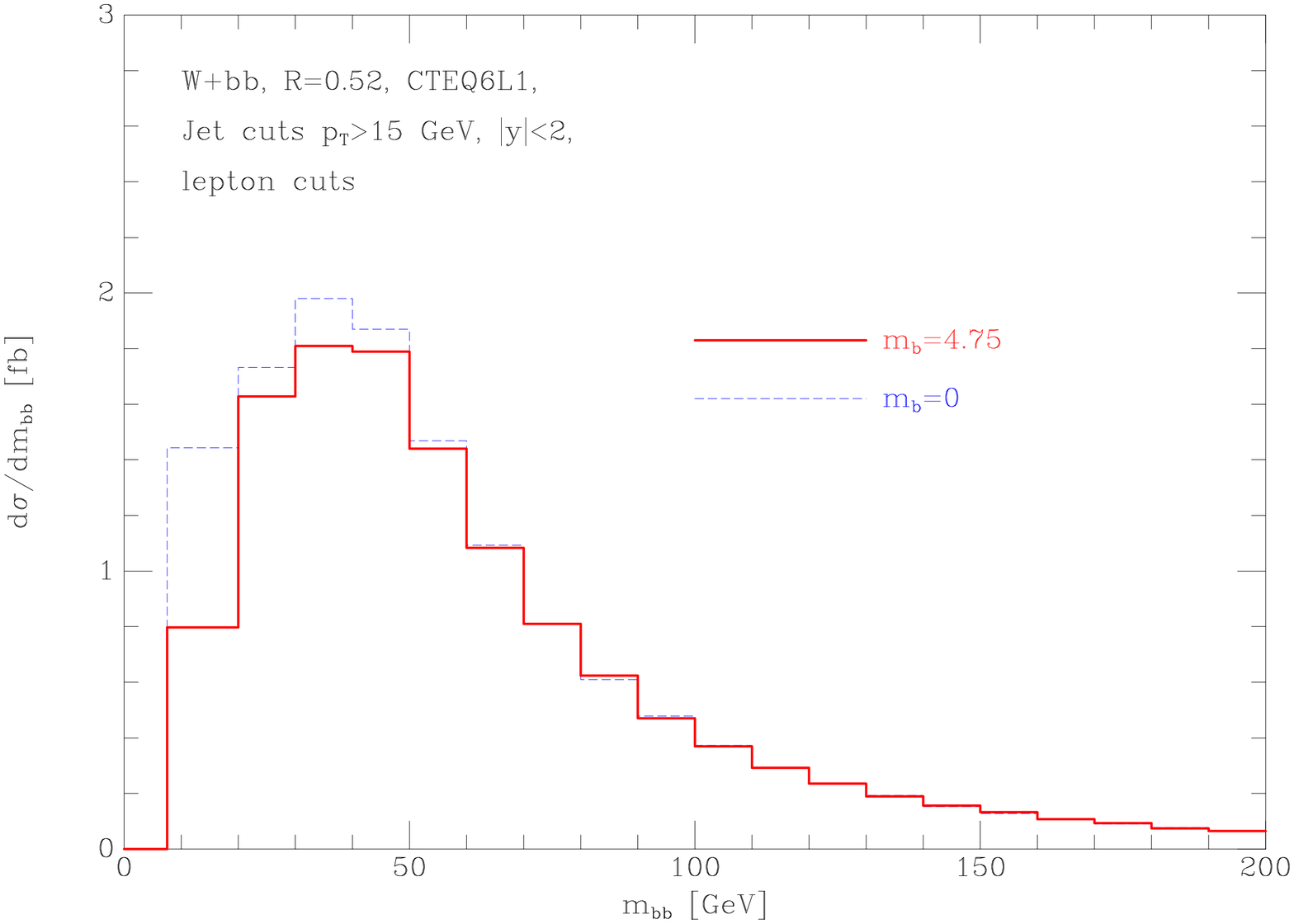}
\end{center}
\caption{ The $b\overline{b}$ invariant mass in $Wb\overline{b}$ events, using
lowest order matrix elements. 
\label{fig:wbb_mass}}
\end{figure}

\begin{figure}
\begin{center}
\includegraphics[scale=0.4]{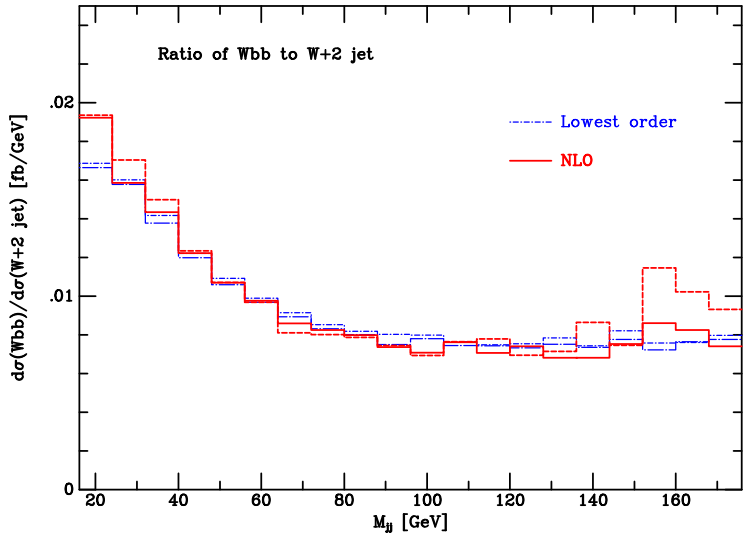}
\end{center}
\caption{ The ratio of the cross section for $Wb\overline{b} $ to $Wjj$ production as a function of the dijet  mass. 
\label{fig:wbb_wjj_ratio}}
\end{figure}

The $b\overline{b}$ pair invariant mass distribution peaks at low values of the
mass due to the gluon propagator. This can be seen in Figure~\ref{fig:wbb_mass},
where we show the lowest order prediction using MCFM. For $W + 2$~jet production,
$t$-channel dijet production dominates so that the dijet  masses will in general
be higher for $Wjj$ than for $Wb\overline{b}$ final states. Indeed,  this is the
case as can be observed in Figure~\ref{fig:wbb_wjj_ratio}, where the ratio of
$Wb\overline{b}/Wjj$ is plotted as a function of the dijet mass. 

The Feynman diagram topologies may give some indication of reasonable values for
the renormalization and factorization scales to use for the matrix element
evaluation at leading order. 
The next-to-leading order result may provide further evidence
to support a particular scale(s) deemed appropriate at  leading order. 

\section{Partonic Cross Sections and Jet Algorithms}

MCFM is a parton-level event generator with at most 3 partons in the final state.
No information is available at the hadron level; thus any jet algorithms must be
applied to the 1 or 2 partons that comprise any of the predicted final state jets.

Thus far in Run II, W+jet cross sections have been measured only with cone jet
algorithms (although the ultimate goal is to also utilize $k_T$ algorithms as
well). There are two options: the Run I legacy algorithm (JetCLU) and the joint
CDF-D0 Run II algorithm (midpoint). The midpoint algorithm is so-named because it
places a seed at the energy-weighted midpoint between two partons, something the
JetClu algorithm does not. The midpoint algorithm also lacks the JetClu feature of
``ratcheting'', where seed towers  initially in the jet cone are added to the
final jet energy, even if the final cone should not nominally include these
towers.~\footnote{No seed tower left behind.} Such an effect is difficult to model
at  the partonic level. The two effects end up being in the opposite direction so
the differences between the two algorithms should be small (5\% or less). One
notable difference between the two algorithms is that the midpoint algorithm is
defined in terms of the transverse momentum ($p_T$) rather than the transverse
energy ($E_T$). Details of the application of the two algorithms to partonic level
cross sections can be found in Ref.~\cite{ratchetpaper}. For historical reasons,
the cross sections generated thus far using MCFM have used a $k_T$ algorithm using
parameters similar to a cone algorithm of radius 0.4; however, this should not
affect any of the conclusions of this paper. Calculations using cone algorithms,
appropriate for direct comparisons to the measured Run 2 cross sections, are the
subject of current study~\cite{inprep}. 

Jet energies measured in the CDF detector have to be corrected for comparison to
theoretical predictions. The level of correction can be tricky for comparison to
calculations at the NLO level. For example, one does not want to correct for the
energy deposited out-of-cone due to perturbative gluon emission, since this is
already accounted for to some level in the theoretical calculation. A correction
should be made, however, for hadronization out-of-cone effects, since these are
not present in the partonic calculation. The average hadronization correction per
jet is on the order of $1$~GeV. 

\section{Scale Dependence}

\begin{figure}
\begin{center}
\includegraphics[scale=0.6]{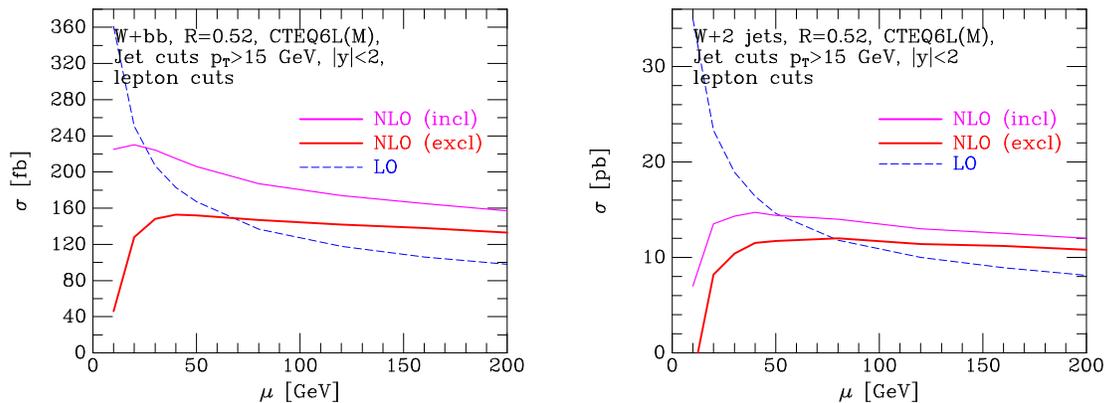}
\end{center}
\caption{ The scale dependence for LO, NLO exclusive and NLO inclusive
$Wb\overline{b}$ and $Wjj$ production, using our usual set of cuts.
The renormalization and factorization scales are equal.
\label{fig:joey}}
\end{figure}

      In Figure~\ref{fig:joey}, the scale dependence for $Wb\overline{b}$ and $Wjj$
production is shown using cuts similar to those used in CDF~\cite{weiming}:
\begin{equation}
R_{cone}=0.4, \quad \Delta R>0.52, 
\quad p_T^{jet}>15~{\rm GeV}, \quad |\eta_{jet}|<2.
\end{equation}
For this calculation, the renormalization and factorization scales have been set
equal. In perturbative QCD the freedom exists to set the two scales separately,
however we have chosen not to do so at this stage. For inclusive cross sections,
there is not the freedom to change the renormalization scale independently at each
vertex,  in contrast to a parton  shower Monte Carlo where the scales at each
vertex may be different. In Section 7, we will consider the impact of
applying separate factorization and renormalization scales.  


      In the calculations performed with MCFM, we distinguish between exclusive
and inclusive production, depending on whether there are exactly two, or two or
more jets that satisfy the kinematic cuts. Contributions to three-jet final states
arise only from the tree level $Wb\overline{b}j(Wjjj)$ diagrams. 

      The leading order cross sections for both processes decrease monotonically
as the renormalization/factorization scale increases. Both the strong coupling
constant $\alpha_s$ and the parton distribution functions (in the relevant
kinematic range) decrease with increasing scale. At NLO, the scale dependence is
reduced for both processes and for both inclusive and exclusive production. The
logarithms that are responsible for the large variations under change of scale at
leading order are exactly cancelled through next-to-leading order; any remaining
scale dependence is at higher order still. 

      At NLO, the cross section typically increases  slightly as the scale
decreases and then at some point peaks and then drops with decreasing scale, due
primarily to the same logarithms that cancel out the scale dependence to NLO. The
exact point at which the maximum of the cross section occurs depends both on the
process under consideration as well as the kinematic cuts, for example the minimum
jet transverse momentum. 

      The scale dependence for $W + 2$~jets seems to be under good control for
both the inclusive and exclusive final states, as long as a scale larger than
$30$~GeV is chosen. For exclusive $Wb\overline{b}$ production, the scale
dependence is also reasonably small for scales larger than $30$~GeV, but a
considerable scale dependence remains for inclusive final states. This is due to
the relatively large number of new channels (with $gq,gg$ initial states)
available for $Wb\overline{b}j$ production at NLO, and the fact that these
processes are only calculated at tree level and thus have a leading order scale
dependence.

\begin{figure}
\begin{center}
\includegraphics[scale=0.6]{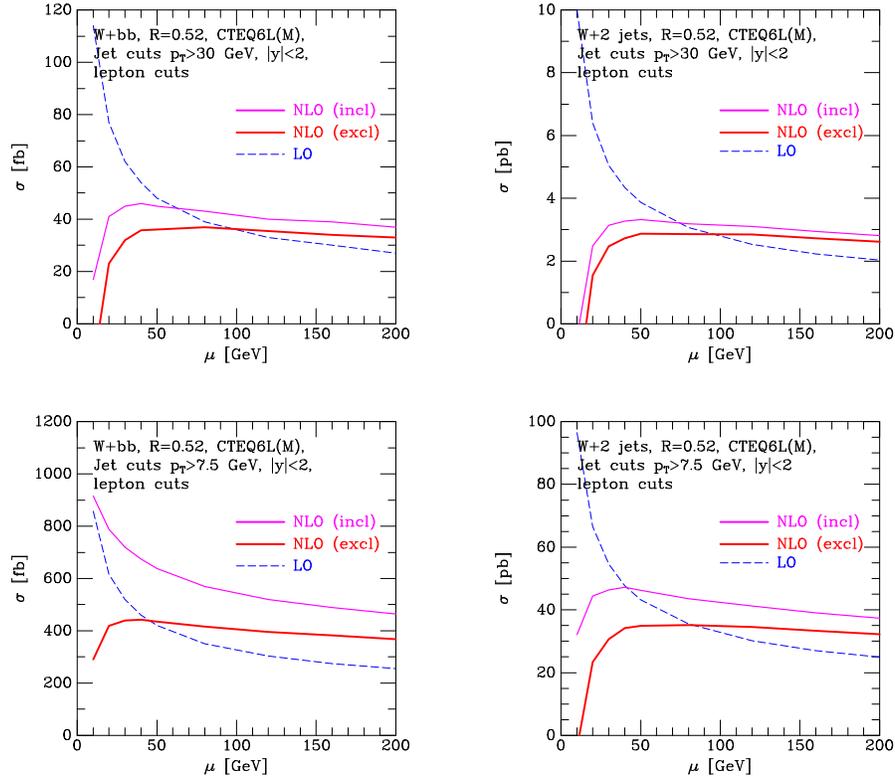}
\end{center}
\caption{ The scale dependence for LO, NLO exclusive and NLO inclusive
$Wb\overline{b}$ and $Wjj$ production when using a higher jet $p_T$ cut
of $30$~GeV (upper plots) and a lower value of $7.5$~GeV (below).
\label{fig:mudep_ptdep}}
\end{figure}

      It is interesting to examine the impact of  changing the kinematic cuts on
the stability of the NLO calculation. In Figure~\ref{fig:mudep_ptdep}, the study
is repeated requiring the jet threshold to be $7.5$~GeV and $30$~GeV. As expected,
the inclusive scale dependence gets worse (better) for the $7.5$~GeV ($30$~GeV)
jet cut, as the $3$-parton final state contribution increases (decreases). Note that
for $Wb\overline{b}$, the scale at which the NLO cross section becomes unstable
moves to higher values as the jet transverse momentum cut increases. 

\begin{figure}
\begin{center}
\includegraphics[scale=0.4,angle=-90]{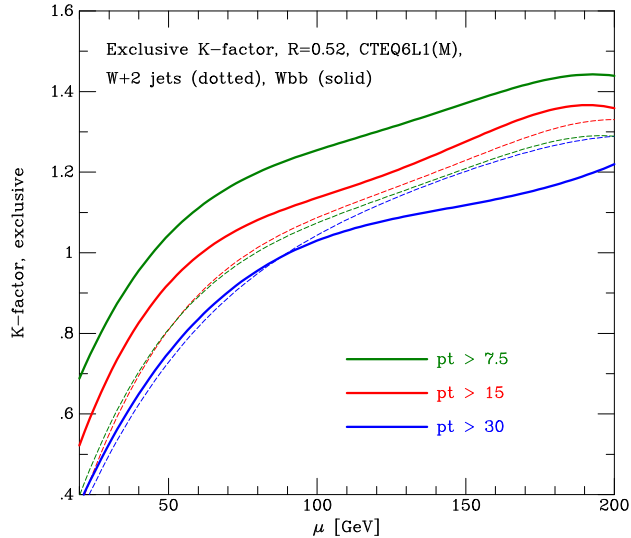}
\end{center}
\caption{ $K$-factors for  $Wb\overline{b}$~(solid curves) and
$Wjj$~(dotted) for exclusive final states for the 3 different jet transverse momentum cuts.
\label{fig:Kfactors}}
\end{figure}

      The $K$-factors (NLO/LO) are shown in Figure~\ref{fig:Kfactors} for 
$Wb\overline{b}$ and $Wjj$ exclusive final states for the three different jet
transverse momentum cuts. The point at which the NLO cross  section equals the LO
cross section (i.e. the $K$-factor is $1$) is relatively insensitive to the jet
transverse momentum cuts for the case of $Wjj$ but systematically moves  out to
higher scale values for $Wb\overline{b}$ production as the jet $p_T$ cut
increases. For a scale of $100$~GeV and for  a jet $p_T$ cut of $15$~GeV, the
$Wb\overline{b}$ $K$-factor is $1.15$ while the $Wjj$ $K$-factor is $1.05$. 

\begin{figure}
\begin{center}
\includegraphics[scale=0.4]{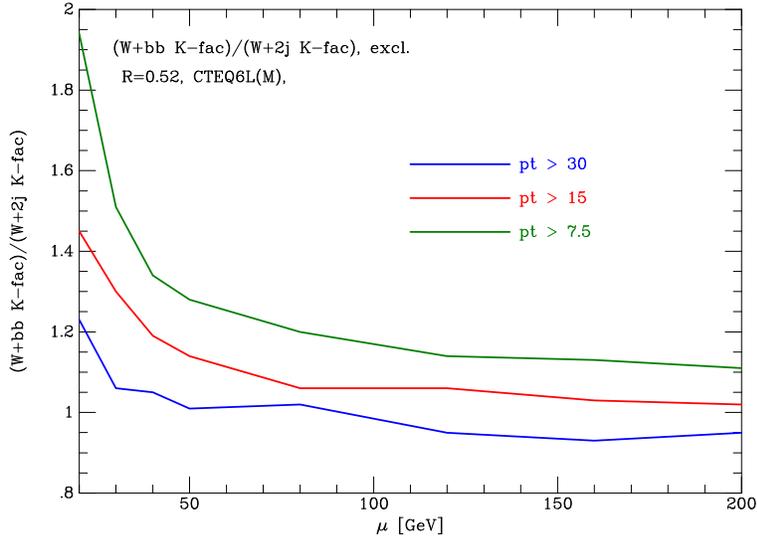}
\end{center}
\caption{ The ratio of the $K$-factors for the two processes, as a
function of the common renormalization and factorization scale $\mu$.
\label{fig:Kf_ratio} }
\end{figure}

      It  is also interesting to plot  the ratio of the $K$-factors of the two
processes ($Wb\overline{b},Wjj$) as a function of the scale. The ratio of
$K$-factors, shown in Figure~\ref{fig:Kf_ratio},  has a strong dependence on the
jet  $p_T$ cut because the $Wb\overline{b}$ $K$-factor does. The ratio is observed
to be relatively constant for scales on the order of $100$~GeV or above. The ratio
is particularly unstable for  low scales. 


\section{Top Physics}

\subsection{Two  Jet Bin}

      One of the most promising channels for searching for $t\overline{t}$ events
is a final state consisting of a high $p_T$ lepton plus missing transverse energy
plus jets. The $W$ boson from one of the tops has decayed into a an electron or
muon and neutrino while  the $W$ from the other top has decayed into two quarks.
Thus, there will be $4$ partons in the final state and one expects $t\overline{t}$
final states to have their largest contributions in the $W + 3,4$~jet bins,
depending on the jet transverse momentum cuts. Lacking NLO calculations of
$Wb\overline{b}jj$ final states, CDF has estimated the number of
non-$t\overline{t}$ events in the $W+$~jets sample by calculating  the theoretical
ratio of the $Wb\overline{b}+ (n-2)$ jet cross section to the $W + n$~jet cross
section ($n=2,3,4$) and then multiplying this ratio by the  observed number  of 
jets in the $W + n$~jet bin. Individually, such  LO cross sections of course have
a large scale dependence and thus a large uncertainty. The hope is that the
$K$-factors and the scale dependence will be similar  for the two processes and
thus the LO ratio will be an accurate estimate of the NLO one.  In CDF, this has
been known as $\it Method~2$. Such assumptions can be explicitly tested in the
$Wb\overline{b}/Wjj$ case. In Figure~\ref{fig:Kf_ratio}, the ratio of the
K-factors for the exclusive processes $Wb\overline{b}$ and $Wjj$ were plotted as a
function of the scale for three different jet $p_T$ cuts. For Method 2 to be
accurate,  the ratio should be  near unity. This is true
($K_{Wb\overline{b}}/K_{Wjj}=1.05$) for a jet $p_T$ cut  of $15$~GeV, but perhaps
a kinematic accident rather than a {\it God-given truth}. It is not
true for for any scale for a $p_T$ cut of $7.5$~GeV nor for small scales 
($20-30$~GeV) for  any jet $p_T$ cut. 

\begin{figure}
\begin{center}
\includegraphics[scale=0.8]{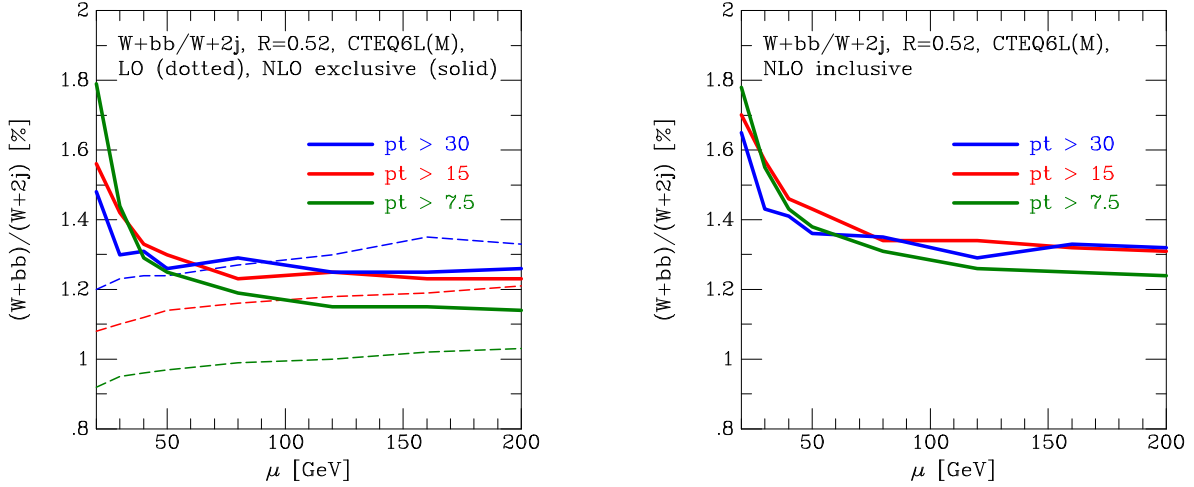}
\end{center}
\caption{ The ratio of the $Wb\overline{b}/Wjj$ cross sections for  exclusive and inclusive production.
\label{fig:wbb_wjj} }
\end{figure}

      The ratio of the $Wb\overline{b}/Wjj$ cross sections is shown in
Figure~\ref{fig:wbb_wjj} for both exclusive and inclusive production. For  both
cases, the ratio is examined as a function of  the minimum cut on the  jet 
transverse momentum.  It  is interesting to note that for low scales, the ratio is
more  stable at LO than at NLO. As noted earlier, this is due to the relatively
large amount  of tree-level three-parton final  states that enter into the
$Wb\overline{b}$ process at NLO. At NLO, the ratio $Wb\overline{b}/Wjj$ is
approximately $1.25$\% for a $15$~GeV jet cut, while  at  LO this ratio is
$1.18$\% (but again very sensitive to the kinematic cuts). For lower scales, the
discrepancy is much more extreme. At LO,  the ratio has a strong dependence on the
jet transverse momentum cut; this dependence is greatly reduced at NLO. A scale of
approximately $100$~GeV is in the region of stability at NLO. For this scale, for
a jet $p_T$ cut of $15$~GeV, the $K$-factor  is also on the order of unity (as
already noted). The inclusive ratio also has the instability at small scales and
is approximately $0.1$ higher than the exclusive case for all scale values. 

The exclusive NLO ratio for $Wb\overline{b}/Wjj$ of $1.25$\% calculated at the scale
of $100$~GeV is larger by a factor of $1.4$ than  the value assumed for the $2$ jet bin
using ALPGEN at LO in the CDF Run II top analysis~\cite{weiming}. However, it is
in good agreement with the value calculated for the $2$ jet bin in Run I, after an
empirical factor of $1.4$, determined from jet data~\cite{run1kfac}, is applied. Thus, this analysis
can be considered as a validation of the factor of $1.4$. Note that the leading
order value of the ratio in the $2$ jet bin in this analysis is larger than the
value observed in ALPGEN. This may be due in part to the $b$ quark being massless
in this analysis, but not in the ALPGEN calculation.


\begin{figure}
\begin{center}
\includegraphics[scale=0.6]{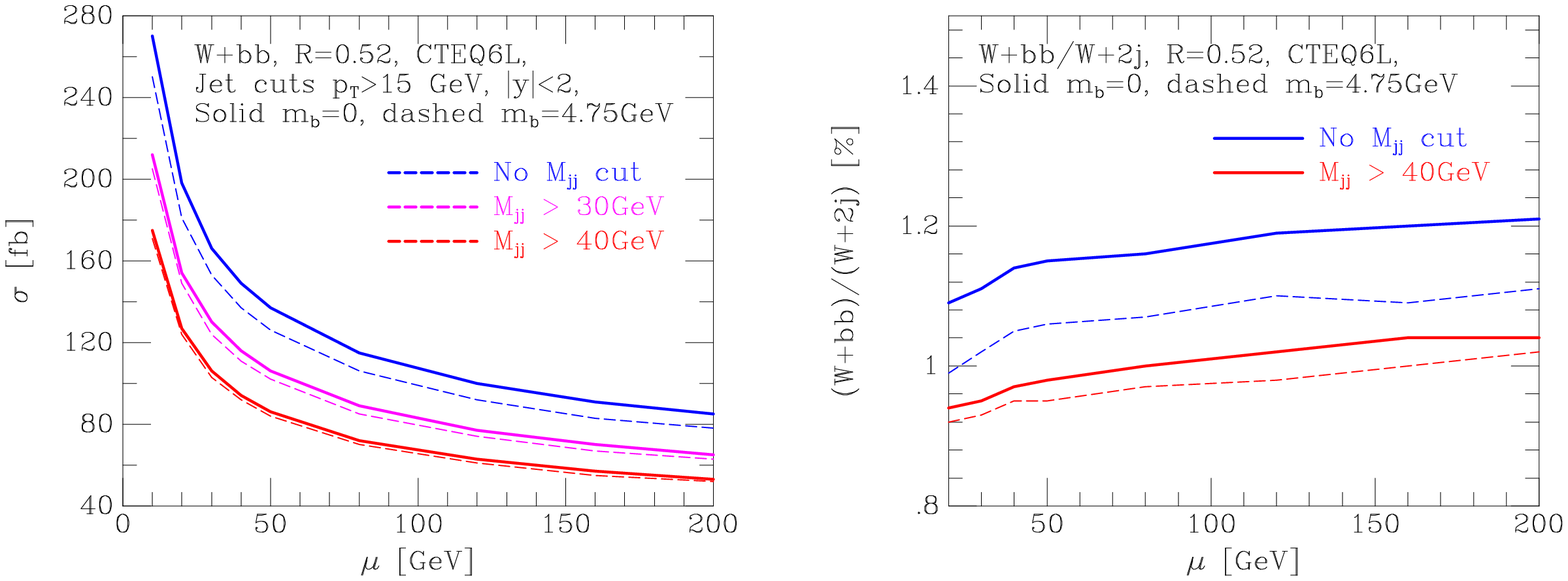}
\end{center}
\caption{ The effect of a dijet mass cut on $Wb\overline{b}$ cross section and 
on the  $Wb\overline{b}/Wjj$ ratio.
\label{fig:mcut} }
\end{figure}

      It was noted earlier that the $b\overline{b}$ mass peaks more strongly at
low mass than does the dijet mass. We also  examine the lowest order
$Wb\overline{b}/Wjj$ ratio requiring $m_{b\overline{b}}$ ($m_{jj}$) be  greater 
than $20$, $30$ and $40$ GeV, as shown in Figure~\ref{fig:mcut}. As expected, the
ratio $Wb\overline{b}/Wjj$ decreases as the $b\overline{b}$ mass cut is increased.
Furthermore, the impact of neglecting the $b$ mass in the calculation decreases
as the $b\overline{b}$ mass cut is increased. 

\subsection{Higher Jet Multiplicity Bins and Kinematical Analyses}

\begin{figure}
\begin{center}
\includegraphics[scale=0.5]{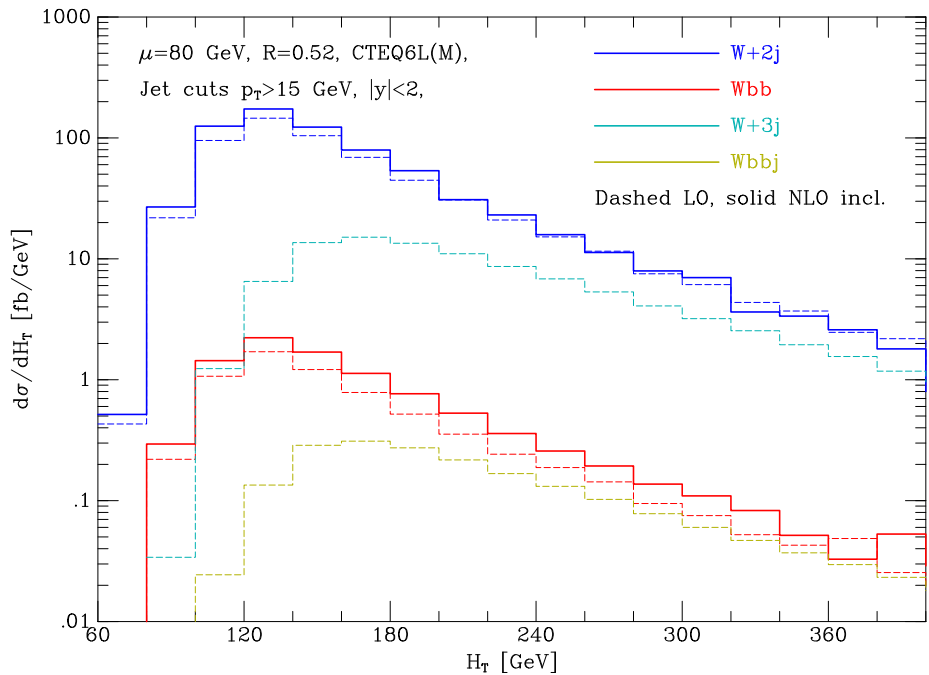}
\end{center}
\caption{ The $H_T$ distributions for $Wb\overline{b}(j)$ (lower curves)
 and $Wjj(j)$ (upper curves).
\label{fig:ht_80} }
\end{figure}

      NLO calculations are not yet available for the $3$ and $4$ jet bins;
however, the inclusive ratio for $Wb\overline{b}/Wjj$ does give us some
information about these higher jet multiplicity states. 
 The topology of $Wb\overline{b} + (n-2)$ jet final states still remains
fundamentally different from $W + n$~jet final states because of the importance of
the $g\rightarrow b\overline{b}$ vertex in the former. Gluon splitting is a
negligible contribution to the $W +$~jets sample.  The strong scale dependence
observed for $Wb\overline{b}+2$~jets at NLO may lead to some wariness about
predictions for the behavior of higher jet multiplicity states. However, as we
have argued, the scale sensitivity results from  the relative importance of the
$Wb\overline{b}$~+jet tree level diagrams; this will not happen at NLO for  the $3$
or $4$ jet bin. In lieu of NLO higher jet multiplicity calculations, we can try to
form observables from suitably inclusive quantities for which the NLO
$Wb\overline{b}$ and $Wjj$ calculations may form an adequate approximation. One
such observable is $H_T$; here, $H_T$ is defined as the sum of the transverse
momenta of  all of the jets, the lepton and the missing $p_T$ in the event.  The
$H_T$ distributions for $Wb\overline{b}(j)$ and $Wjj(j)$ are plotted in
Figure~\ref{fig:ht_80}. The (LO) distributions for $Wb\overline{b}j$ and $Wjjj$
naturally peak at  at a higher value of $H_T$ due to the requirement of an
additional jet.

\begin{figure}
\begin{center}
\includegraphics[scale=0.5]{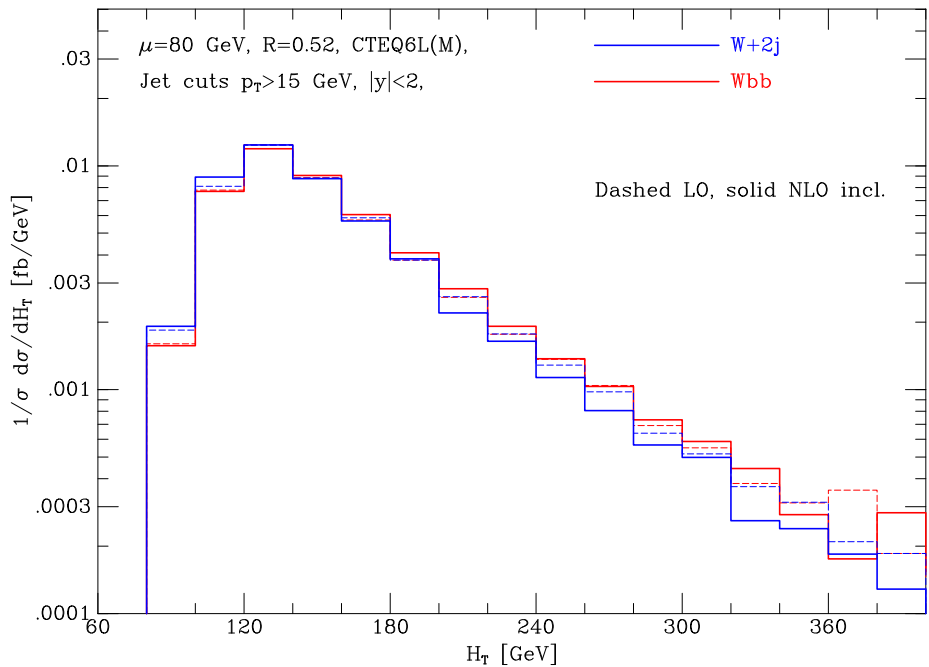}
\end{center}
\caption{ The $H_T$ distributions for $Wb\overline{b}(j)$ and $Wjj(j)$, normalized to the same area.
\label{fig:ht_80_norm}}
\end{figure}

For $Wjj$ the $H_T$ distributions are similar at LO and NLO. The $Wb\overline{b}$
distribution is steeper at LO than at NLO. In Figure~\ref{fig:ht_80_norm}, the
cross sections for $Wb\overline{b}$ and $Wjj$ are normalized for shape
comparison.  There it can be clearly observed that, although, the $Wb\overline{b}$
and $Wjj$ shapes are similar at LO, the $Wjj$ distribution is steeper than that of
$Wb\overline{b}$ at NLO. Any assumption that the shapes of the two processes are
similar (as for example in some kinematical analyses) can be dangerous.

\begin{figure}
\begin{center}
\includegraphics[scale=0.5]{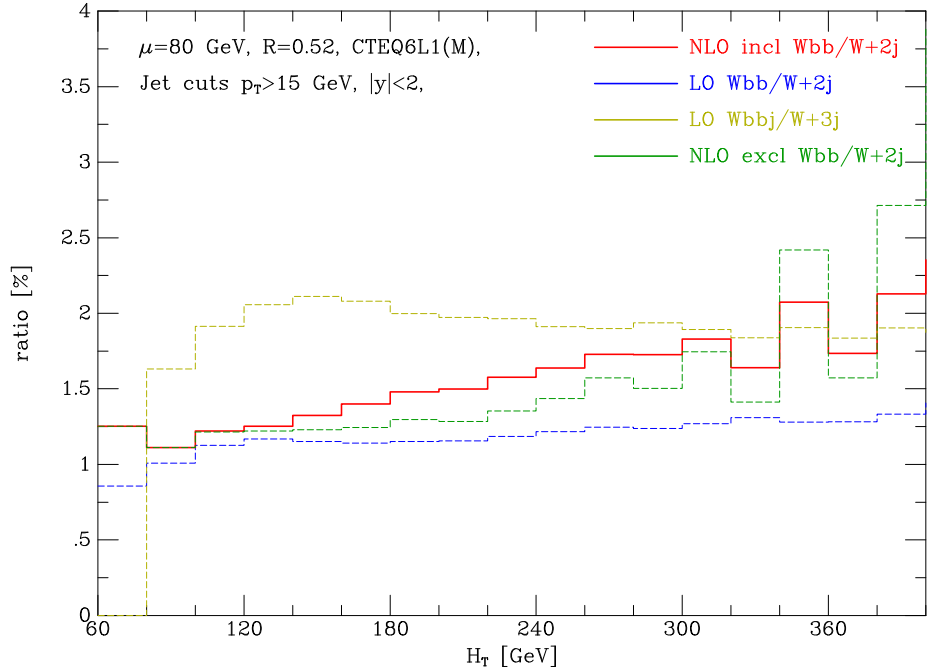}
\end{center}
\caption{ The ratio of the $H_T$ distributions for $Wb\overline{b}(j)$ and $Wjj(j)$.
\label{fig:ht_80_ratio}}
\end{figure}

In Figure~\ref{fig:ht_80_ratio}, the ratio of $Wb\overline{b}$ to $Wjj$,  at LO
and NLO (both inclusive and exclusive), as well as the (LO) ratio of
$Wb\overline{b}j$ to $Wjjj$ is plotted as a function of $H_T$. Here, the
similarity of the $Wb\overline{b}$ and $Wjj$  distributions at LO can be clearly
observed, as well as the fact that the NLO ratio of $Wb\overline{b}$ to $Wjj$
increases by more than a factor of two over the $H_T$ range of the plot (which is
approximately the same as  the $H_T$ range used in the top analysis).  Similar
results to those above have also  been obtained when replacing $H_T$ by the scalar
sum of the jet transverse energies.

\begin{figure}
\begin{center}
\includegraphics[scale=0.5]{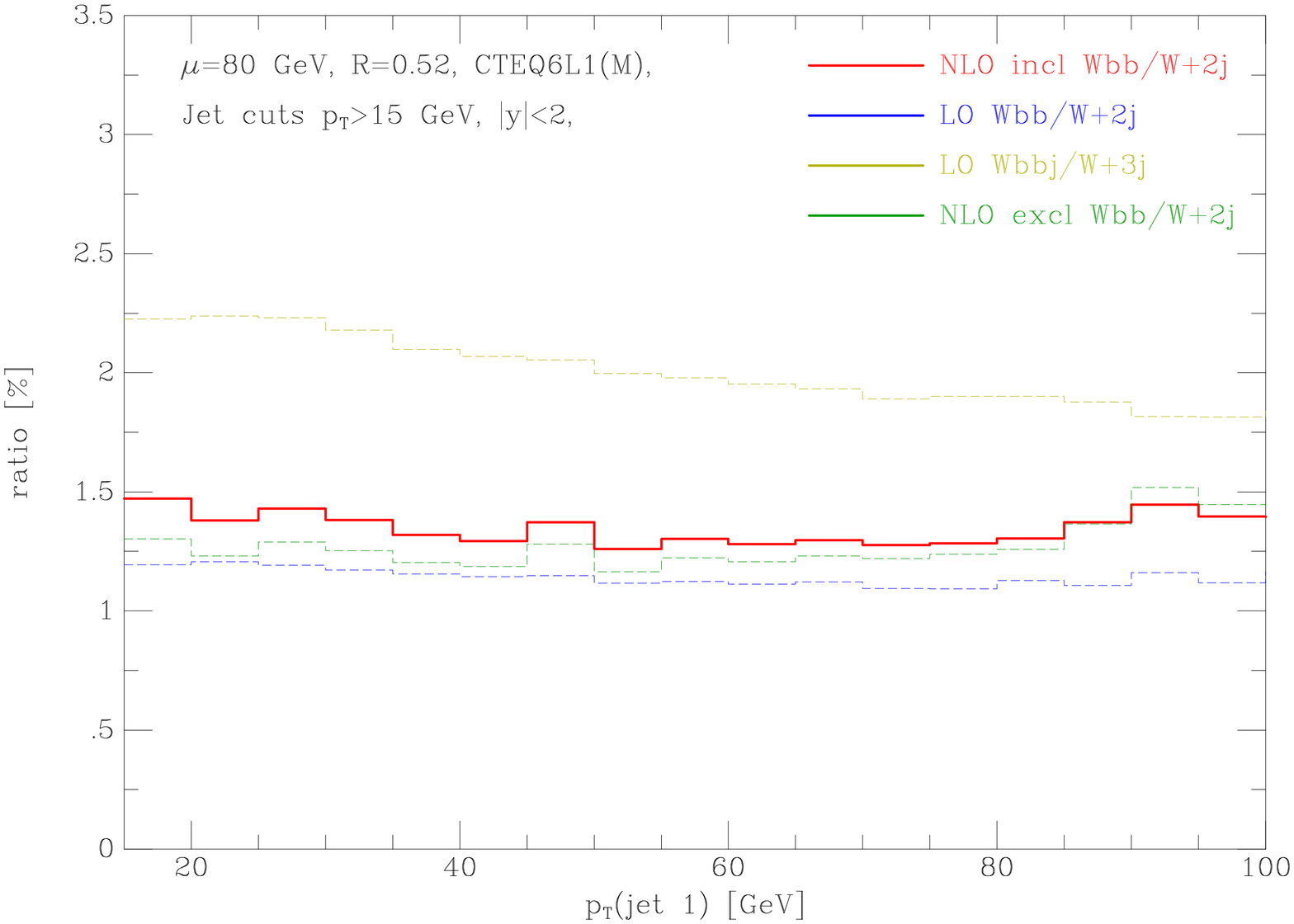}
\end{center}
\caption{ The $p_{Tjet 1}$ distributions for $Wb\overline{b}(j)$ and $Wjj(j)$.
\label{fig:pt1jet_80}}
\end{figure}

Given the shape differences between LO and NLO observed for global variables such
as these, it is important to discover if the effects are the
same for less inclusive observables. In Fig.~\ref{fig:pt1jet_80}
the ratio of $Wb\overline{b}$ to $Wjj$ is plotted as a
function of the lead jet $p_T$.
In contrast to the previous cases, the ratio is flat at NLO for both inclusive and
exclusive final states. Thus, this is an example of a variable which
appears to be {\it safe} with regard to a change of shape at NLO. 

\begin{figure}
\begin{center}
\includegraphics[scale=0.5]{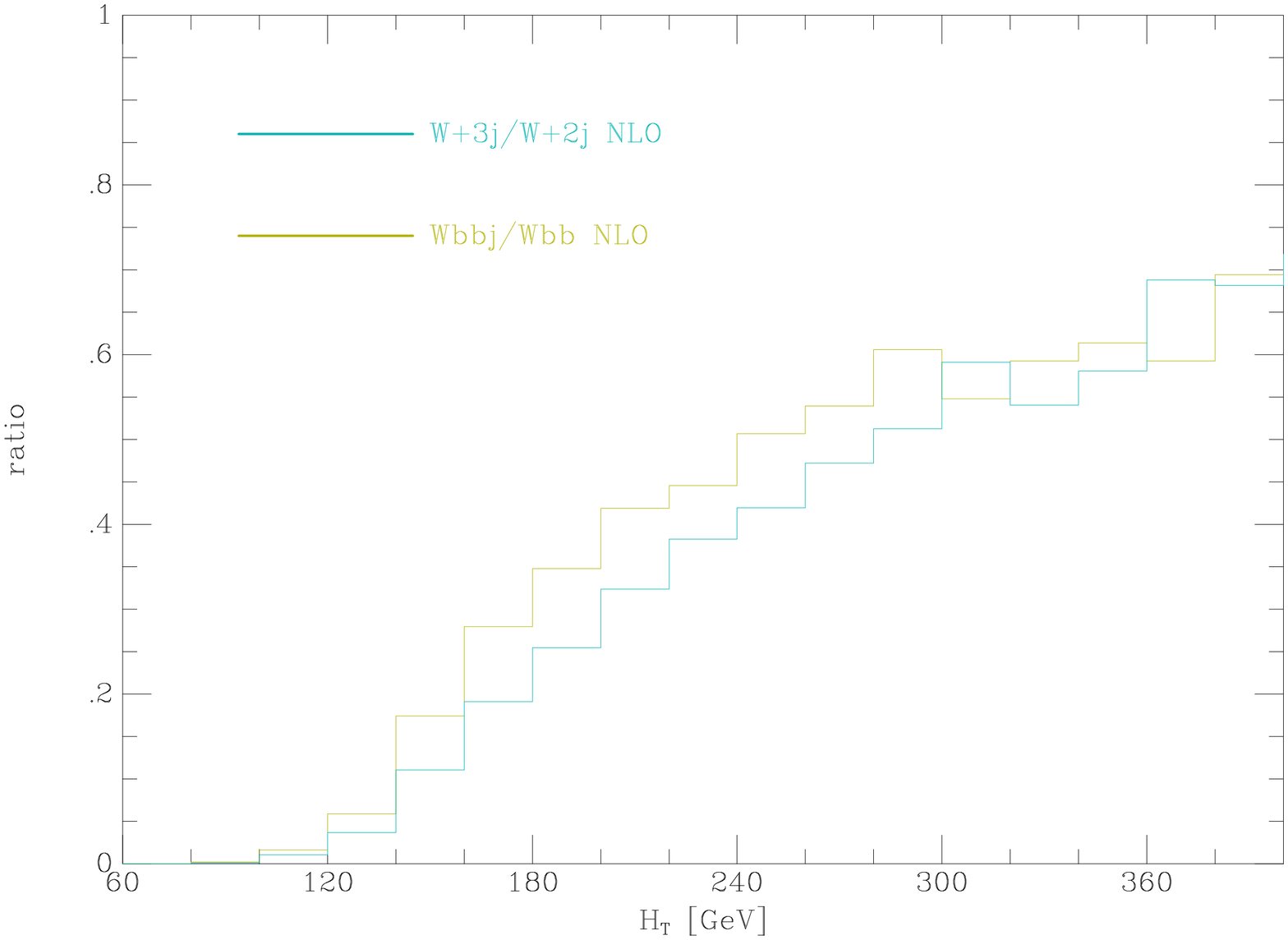}
\end{center}
\caption{ The relative contributions to the NLO $H_T$ cross section of the $Wb\overline{b} (Wjj)$ and $Wb\overline{b}j (Wjjj)$ subprocesses.
\label{fig:ht_rel}}
\end{figure}

\begin{figure}
\begin{center}
\includegraphics[scale=0.5,angle=-90]{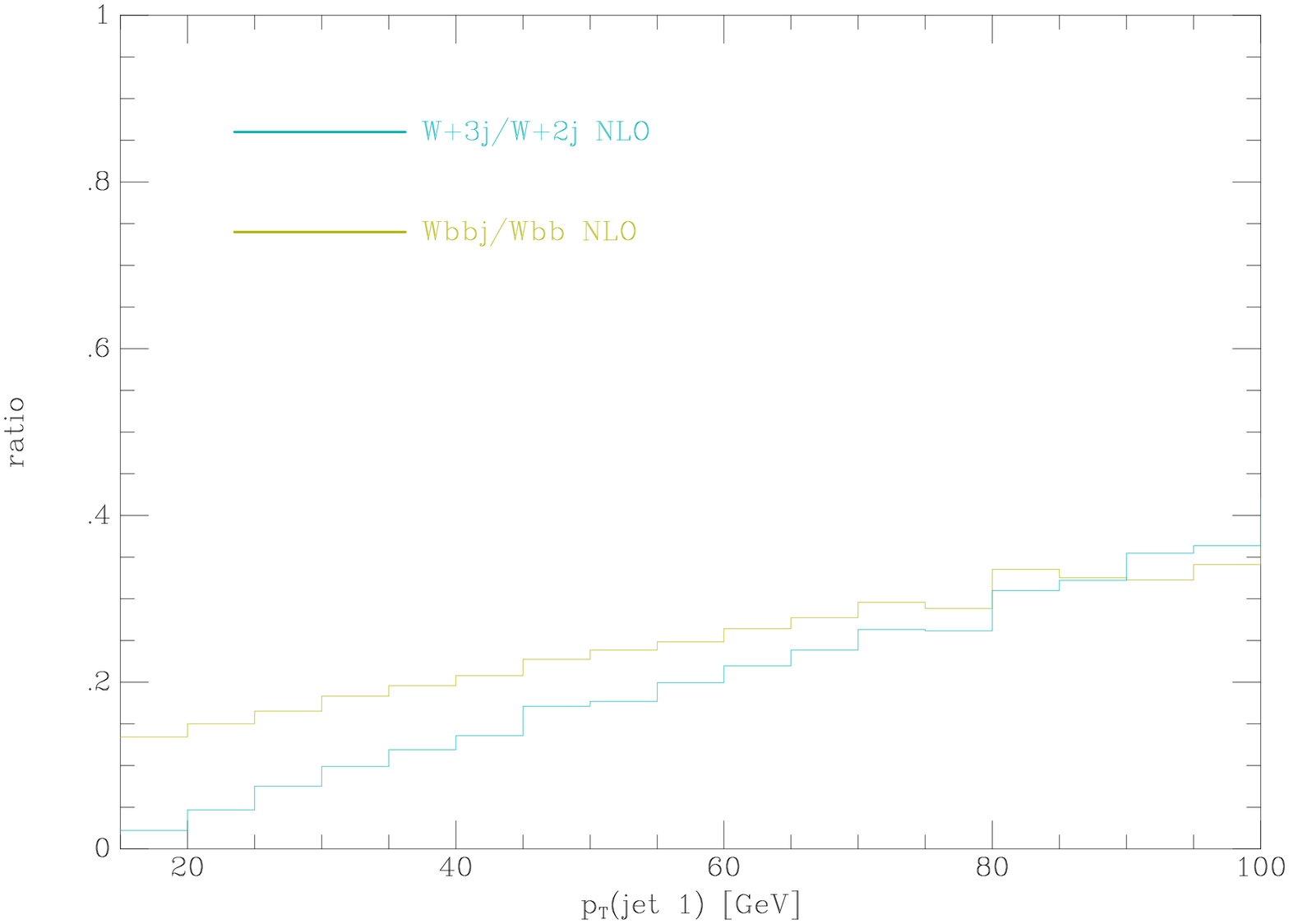}
\end{center}
\caption{ The relative contributions to the largest jet $p_T$ cross section of the $Wb\overline{b} (Wjj)$ and $Wb\overline{b}j (Wjjj)$ subprocesses. 
\label{fig:pt_rel}}
\end{figure}

In Figs.~\ref{fig:ht_rel}~and~\ref{fig:pt_rel}  we show the relative
contributions to the NLO $H_T$ and largest jet $p_T$ cross sections of the
$Wb\overline{b} (Wjj)$ and $Wb\overline{b}j (Wjjj)$ subprocesses. In the NLO
inclusive results, the contribution to the $H_T$ distribution for
$Wb\overline{b}j$ and  $Wjjj$ events is negligible at small $H_T$ and dominant at
large $H_T$, leading to the change in shape seen in Fig.~\ref{fig:ht_80_ratio}.
This can be contrasted with the extra jet contribution to the largest jet $p_T$
distribution, which is never dominant over the range shown. 

\section{Two Scale Studies}

\begin{figure}
\begin{center}
\includegraphics[scale=0.5]{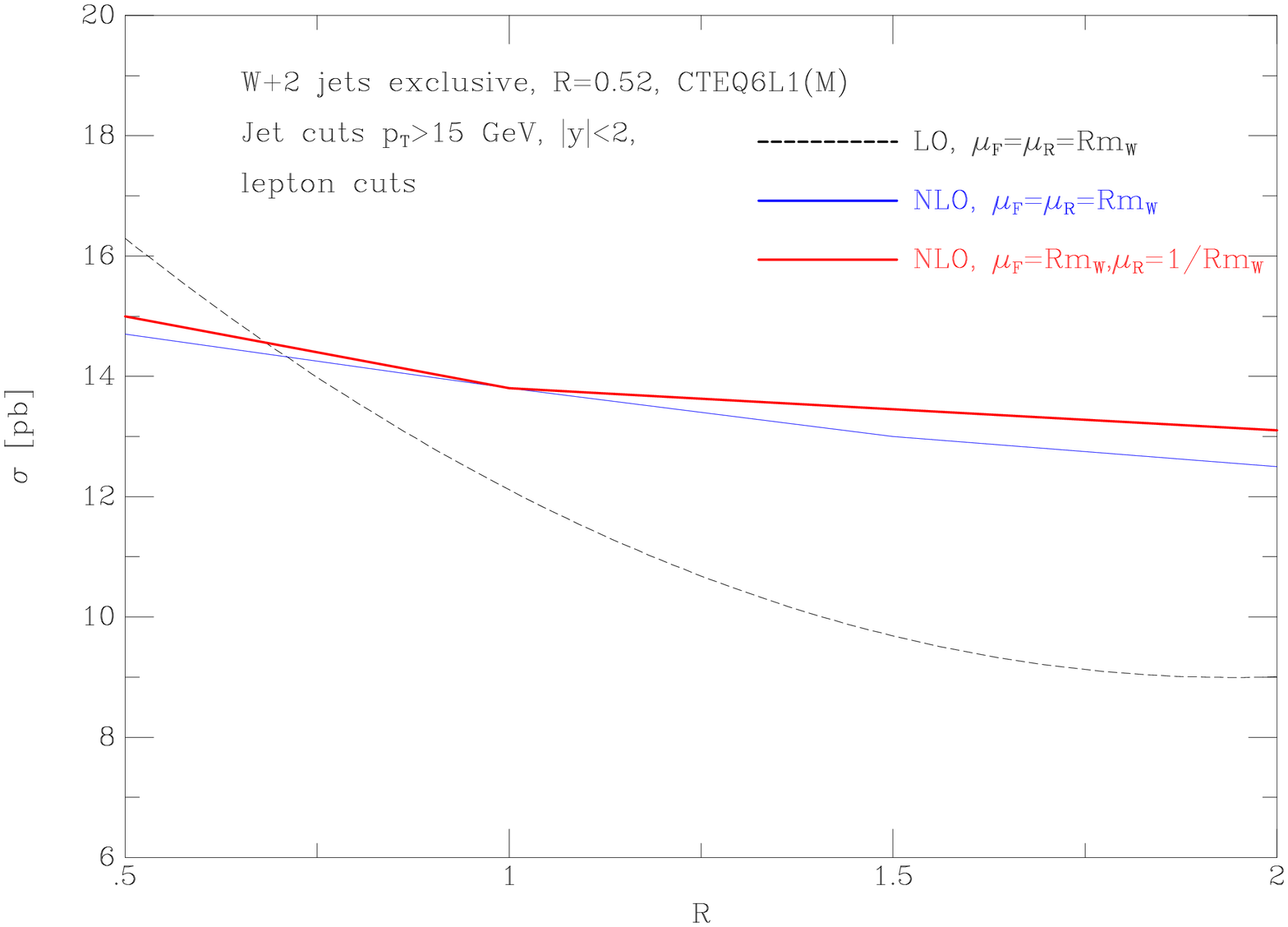}
\end{center}
\caption{The scale dependence of the $Wjj$ exclusive cross-section,
using our usual choice of varying renormalization and factorization
scales together ($\mu_F=\mu_R=R \, M_W$, for $1/2<R<2$), as well as the
choice of varying them in opposite directions
($\mu_F=R \, M_W, \mu_R=1/R \, M_W$). 
\label{fig:mu2dep_W2}}
\end{figure}

\begin{figure}
\begin{center}
\includegraphics[scale=0.5]{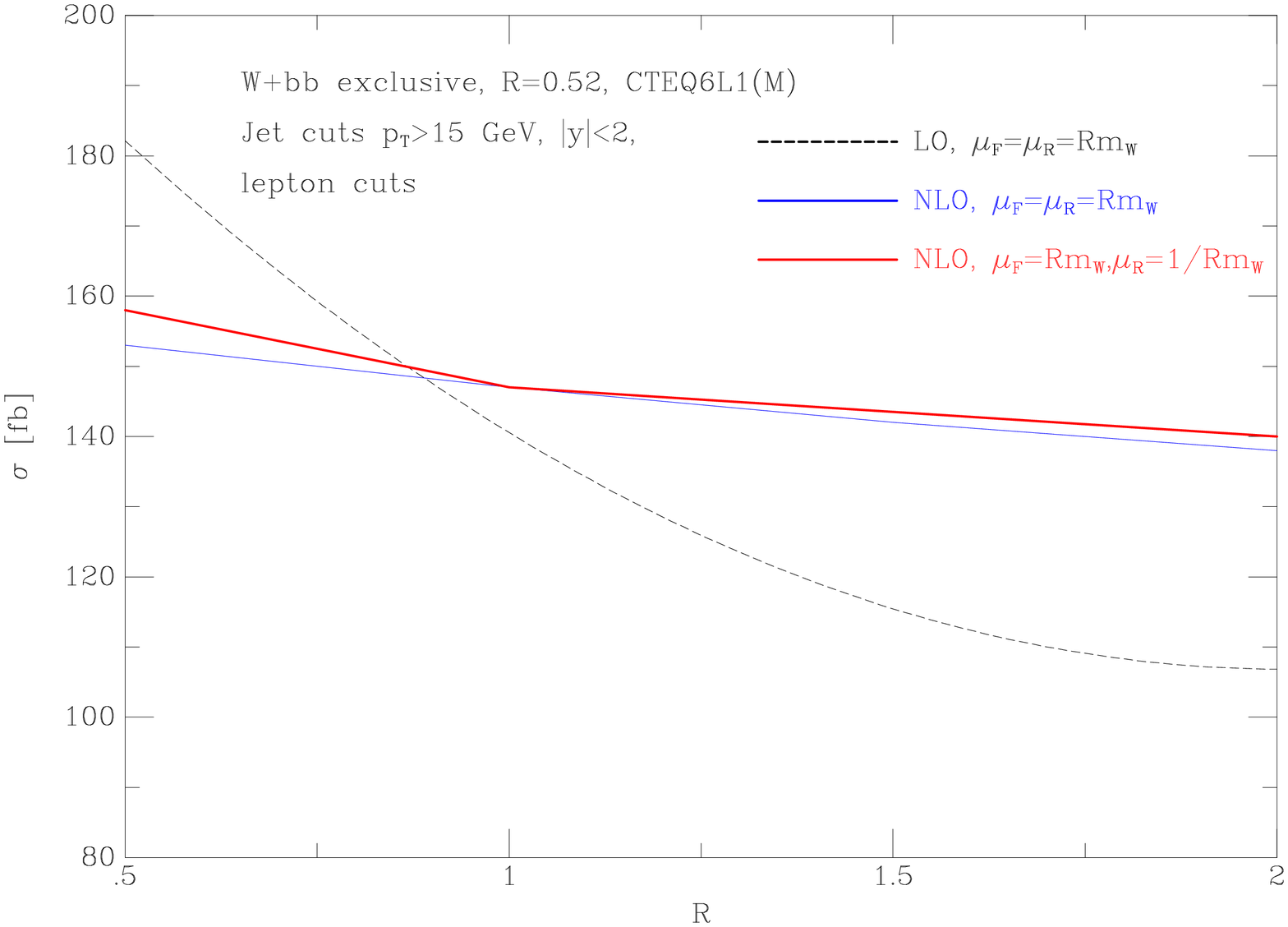}
\end{center}
\caption{The scale dependence of the $Wb{\bar b}$ exclusive cross-section,
using our usual choice of varying renormalization and factorization
scales together ($\mu_F=\mu_R=R \, M_W$, for $1/2<R<2$), as well as the
choice of varying them in opposite directions
($\mu_F=R \, M_W, \mu_R=1/R \, M_W$).
\label{fig:mu2dep_Wbb}}
\end{figure}

Thus far, we have set the renormalization and factorization scales to the same
value. A version of MCFM has been modified to allow the separation of the two
scales. In Figure~\ref{fig:mu2dep_W2}, the $Wjj$ cross section has been plotted with
the two scales set  the same  (both at LO and NLO) as well as for one scale
pre-factor being set to the inverse of the other (at NLO). A similar plot is shown
in Figure~\ref{fig:mu2dep_Wbb}  for the case of $Wb\overline{b}$. As expected, the
scale dependence is large at leading order; at NLO the cross section is reasonably
stable for all of the scale choices plotted. Thus, the ratio of the two cross
sections used in Method 2 should also be stable at NLO for the case of unequal
renormalization and factorization scales and there is no anomalous enhancement of
the $Wb\overline{b}$ cross section at low renormalization scales. 

\section{Other Theoretical Uncertainties}

      One measure of possible NNLO corrections  is supplied by the scale
dependence that remains at NLO. This was shown in Figures 4-9  where it was
observed that the scale sensitivity is under control for choice of scale larger
than $50$~GeV or so  and for reasonably large jet $E_T$ cuts. The ratio can 
also be sensitive to the choice of  parton distribution functions (PDF's). We have
used the $40$ error PDF's supplied with the CTEQ6M central fit PDF to estimate the
contribution  of PDF uncertainty to the ratio of  $Wb\overline{b}$ to $Wjj$
production. As  noted earlier, different initial states contribute to the two
different processes, so that PDF uncertainties will not necessarily cancel. This
is the first time (that we know of) that such a study has been carried out for
$Wb\overline{b}/Wjj$ final states. In order to more easily calculate the PDF
uncertainties, we  have used the LHAPDF interface provided in the most recent
version of MCFM~\cite{LHAPDF}. With this version, the parton
luminosities for each PDF member can be calculated very quickly at each Monte
Carlo integration point. Thus, the cross sections using all members of a given PDF
set can be calculated in one Monte Carlo run. The integration is weighted by the
central PDF luminosity and the result for each error set is recorded separately. 

\begin{figure}
\begin{center}
\includegraphics[scale=0.5]{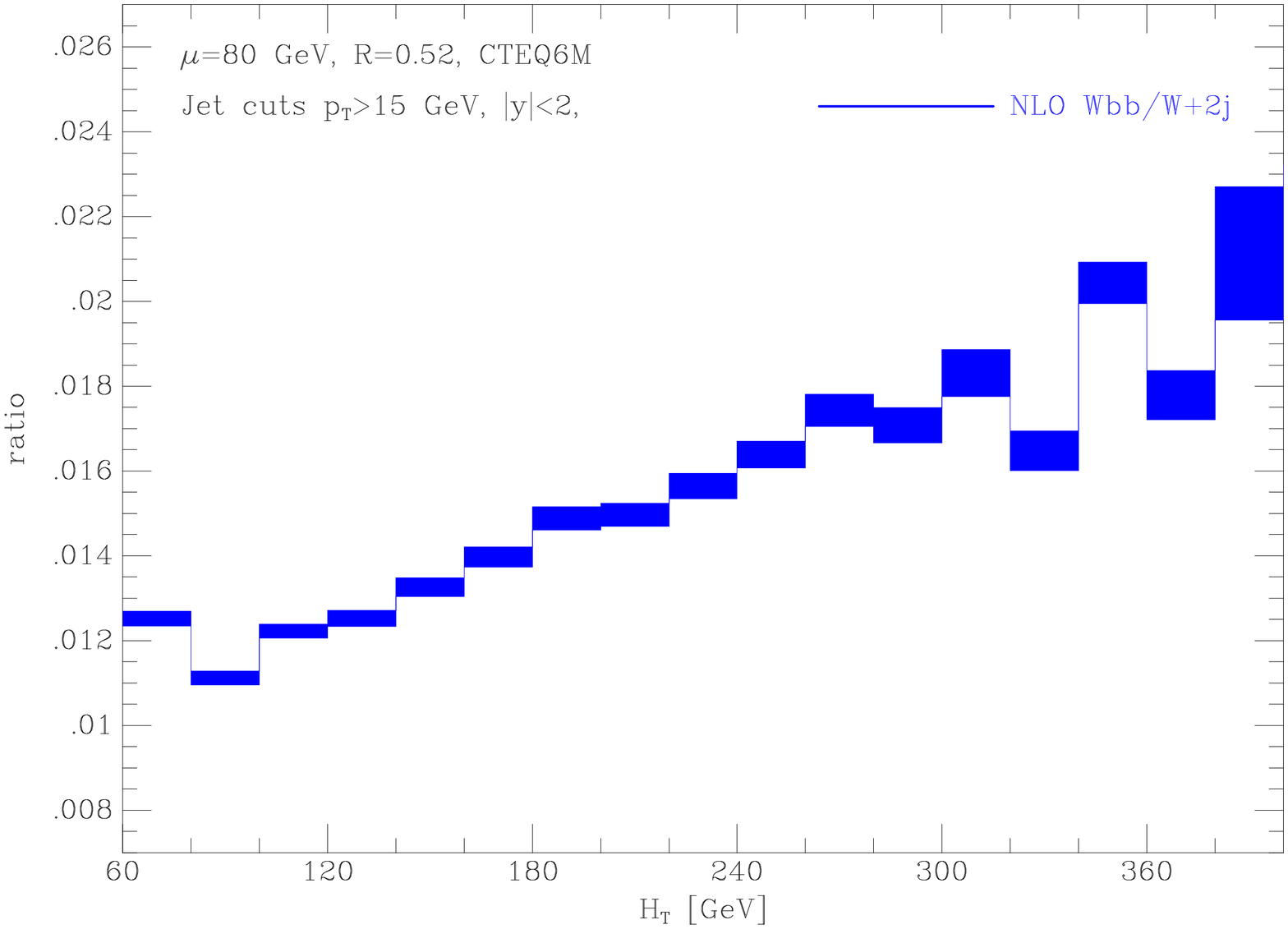}
\end{center}
\caption{ The PDF uncertainty  for the ratio of  $Wb\overline{b}$ and
$Wjj$,  plotted as a function of $H_T$, calculated using the
CTEQ6 error PDF set.
\label{fig:ht_80_ratio_errs}}
\end{figure}

\begin{figure}
\begin{center}
\includegraphics[scale=0.5]{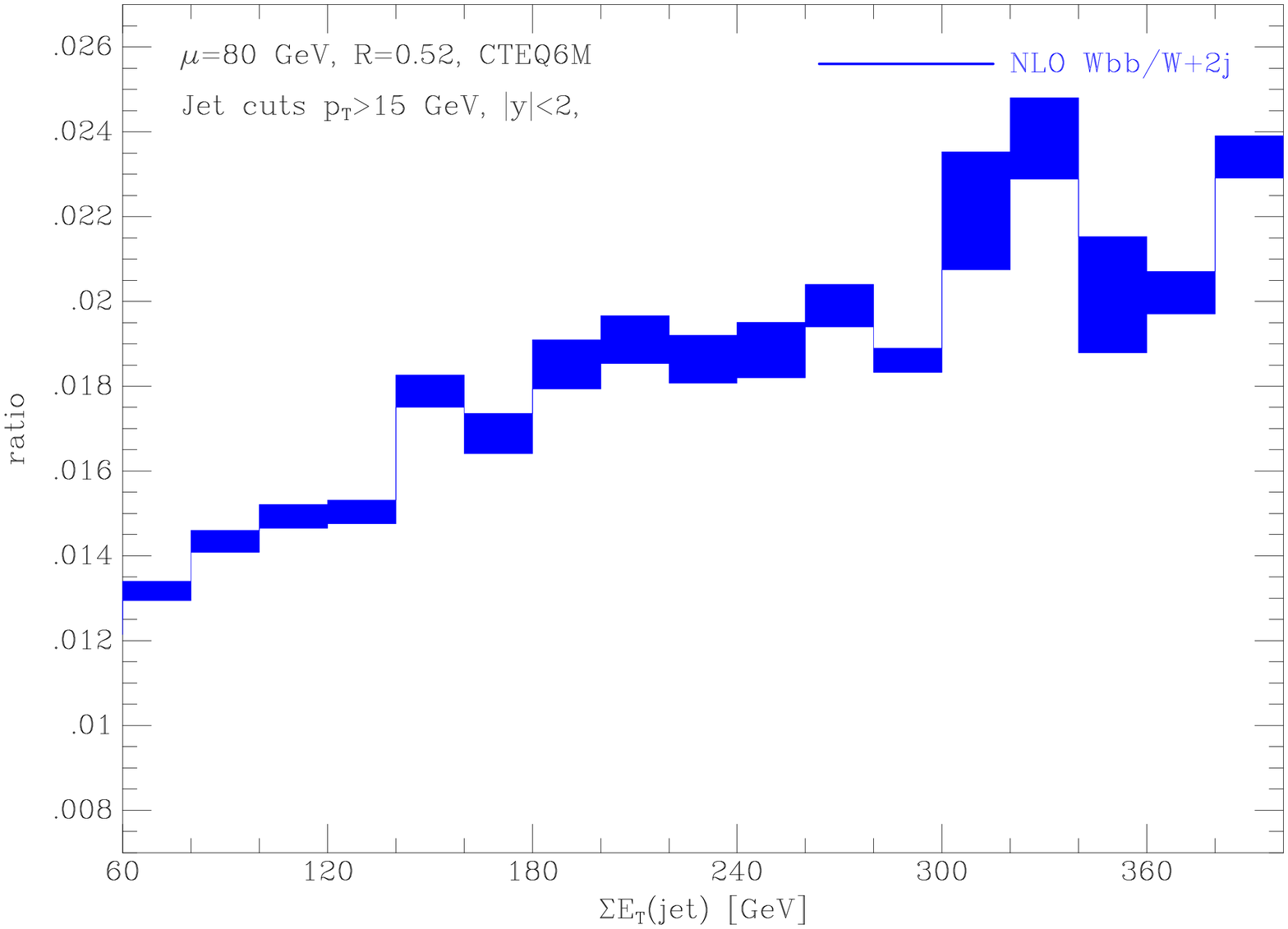}
\end{center}
\caption{ The PDF uncertainty  for the ratio of  $Wb\overline{b}$ and
$Wjj$,  plotted as a function of the sum of the jet transverse energies,
calculated using the CTEQ6 error PDF set.
\label{fig:pt_80_ratio_errs}}
\end{figure}

The PDF uncertainties thus calculated are shown in
Figs.~\ref{fig:ht_80_ratio_errs} and~\ref{fig:pt_80_ratio_errs} for the $H_T$ and
largest jet $p_T$ variables. The PDF uncertainties are reasonably small. 

\section{Conclusions and Future Plans}

Method 2 has proven to be a useful tool for CDF analyses both in Run 1 and in Run
2. In this paper, we have shown that Method 2 holds its validity at NLO for
non-inclusive variables, for jet $E_T$ cuts of the order of $15$~GeV or greater,
and for renormalization/factorization scales of the order of the $W$ mass. 

In an ideal world, we would have available NLO calculations of $Wjjj$, $Wjjjj$,
$Wb\overline{b}j$ and $Wb\overline{b}jj$. Since their availability within the next
year is unlikely, we will have to continue to rely upon LO predictions of these
final states. In a future paper~\cite{inprep}, we will attempt to use the NLO
processes within MCFM to directly compare MCFM predictions to (1) $Wjj$ and
$Wb\overline{b}$ observables in CDF Run 2 data and (2) to enhanced LO predictions
(for example using the CKKW scheme). Good agreement with the latter for $Wjj$ and
$Wb\overline{b}$ final states may give some confidence in the extrapolation to
higher jet multiplicity final states.~\footnote{We will also consider some of the
subtleties generated by comparing parton level calculations to data cross sections
measured with a relatively small ($0.4$) cone size.} In particular, it is believed 
that some explicit higher order corrections are incorporated into the Sudakov form
factors through the use of the CKKW procedure. We can try to explicitly test
this. 

\begin{acknowledgments}
We would like to thank the Fermilab Computing Division for computer
time on the general purpose farm.
This work was supported in part by the U.S. Department of Energy
under Contract No. W-31-109-ENG-38.
\end{acknowledgments}










\end{document}